# An Efficient Inter Carrier Interference Cancellation Schemes for OFDM Systems

B.Sathish Kumar        K.R.Shankar Kumar        R.Radhakrishnan

*Department of Electronics and Communication Engineering*
*Sri Ramakrishna Engineering College*
*Coimbatore, India.*

*Abstract*— **Orthogonal Frequency Division Multiplexing (OFDM) has recently been used widely in wireless communication systems. OFDM is very effective in combating inter-symbol interference and can achieve high data rate in frequency selective channel. For OFDM communication systems, the frequency offsets in mobile radio channels distort the orthogonality between subcarriers resulting in Inter Carrier Interference (ICI). ICI causes power leakage among subcarriers thus degrading the system performance. A well-known problem of OFDM is its sensitivity to frequency offset between the transmitted and received carrier frequencies. There are two deleterious effects caused by frequency offset one is the reduction of signal amplitude in the output of the filters matched to each of the carriers and the second is introduction of ICI from the other carriers. This research work investigates three effective methods for combating the effects of ICI: ICI Self Cancellation (SC), Maximum Likelihood (ML) estimation, and Extended Kalman Filter (EKF) method. These three methods are compared in terms of bit error rate performance and bandwidth efficiency. Through simulations, it is shown that the three techniques are effective in mitigating the modulation schemes, the ML and EKF methods perform better than the SC method.**

## I. INTRODUCTION

OFDM is emerging as the preferred modulation scheme in modern high data rate wireless communication systems. OFDM has been adopted in the European digital audio and video broadcast radio system and is being investigated for broadband indoor wireless communications. Standards such as HIPERLAN2 (High Performance Local Area Network) and IEEE 802.11a and IEEE 802.11b have emerged to support IP-based services. Such systems are based on OFDM and are designed to operate in the 5 GHz band. OFDM is a special case of multi-carrier modulation. Multi-carrier modulation is the concept of splitting a signal into a number of signals, modulating each of these new signals to several frequency channels, and combining the data received on the multiple channels at the receiver. In OFDM, the multiple frequency channels, known as sub-carriers, are orthogonal to each other. One of the principal advantages of OFDM is its utility for transmission at very nearly optimum performance in unequalized channels and in multipath channels.

In this paper, the effects of ICI have been analyzed and three solutions to combat ICI have been presented. The first method is a self-cancellation scheme[1], in which redundant data is transmitted onto adjacent sub-carriers such that the ICI between adjacent sub-carriers cancels out at the receiver. The other two techniques, maximum likelihood (ML) estimation and the extended Kalman filter (EKF) method, statistically estimate the frequency offset and correct the offset [7],using the estimated value at the receiver. The works presented in this paper concentrate on a quantitative ICI power analysis of the ICI cancellation scheme, which has not been studied previously. The average carrier-to-interference power ratio (CIR) is used as the ICI level indicator, and a theoretical CIR expression is derived for the proposed scheme.

## II. IDEALIZED SYSTEM MODEL

The Fig. 1 describes a simple idealized OFDM system model suitable for a time-invariant AWGN channel.

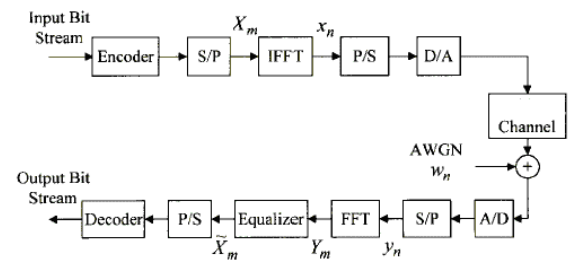

Fig. 1.Idealized OFDM System Model

In an OFDM system, at the transmitter part, a high data-rate input bit stream b[n] is converted into N parallel bit streams each with symbol period Ts through a serial-to-parallel buffer. When the parallel symbol streams are generated, each stream would be modulated and carried over at different center frequencies. The sub-carriers are spaced by 1/NTs in frequency, thus they are orthogonal over the interval (0, Ts). Then, the N symbols are mapped to bins of an Inverse Fast Fourier Transform (IFFT). These IFFT [11] bins correspond to the orthogonal sub-carriers in the OFDM symbol. Therefore, the OFDM symbol can be expressed as

$$X(n) = \frac{1}{N} \sum_{m=0}^{N-1} X_m \, e^{\frac{j2\pi nm}{N}} \qquad (1)$$

where the $X_m$ are the base band symbols on each sub carrier. Then, the $X(i)$ points are converted into a time domain sequence $x(i)$ via an IFFT operation and a parallel to serial conversion. The digital-to-analog (D/A) converter then creates an analog time-domain signal which is transmitted through the channel. At the receiver, the signal is converted back to a discrete N point sequence $y(n)$, corresponding to each subcarrier. This discrete signal is demodulated using an N-point Fast Fourier Transform (FFT) operation at the receiver. The demodulated symbol stream is given by

$$Y(m) = \sum_{n=0}^{N-1} y(n) e^{\frac{-j2\pi nm}{N}} + W(m)k \qquad (2)$$

where w (m) corresponds to the FFT of the samples of w (n), which is the time invariant Additive White Gaussian Noise (AWGN) introduced in the channel Then, the signal is down converted and transformed into a digital sequence after it passes an Analog-to-Digital Converter (ADC). The following step is to pass the remaining T_D samples through a parallel-to-serial converter and to compute N-point FFT. The resulting Yi complex points are the complex baseband representation of the N modulated sub carriers. As the broadband channel has been decomposed into N parallel sub channels, each sub channel needs an equalizer (usually a 1-tap equalizer) in order to compensate the gain and phase introduced by the channel at the sub channel's frequency. These blocks are called Frequency Domain Equalizers (FEQ).Therefore the groups of bits that has been placed on the subcarriers at the transmitter are recovered at the receiver as well as the high data-rate sequence.

## III. ICI SELF CANCELLATION SCHEME

### A. Self-Cancellation

ICI self-cancellation is a scheme that was introduced by Yuping Zhao and Sven-Gustav Häggman[1] in to combat and suppress ICI in OFDM. The main idea is to modulate the input data symbol onto a group of subcarriers with predefined coefficients such that the generated ICI signals within that group cancel each other, hence the name self- cancellation.

### 1) Cancellation Method

In self cancellation scheme the main idea is to modulate the input data symbol on to a group of sub carriers with predefined self coefficients such that the generated ICI signals within the group cancel each other. The data pair $(X, -X)$ is modulated on to two adjacent subcarriers $(l, l+1)$. The ICI signals generated by the subcarrier l will be cancelled out significantly by the ICI generated by the subcarrier $l+1$. The signal data redundancy makes it possible to improve the system performance at the receiver side. In considering a further reduction of ICI, the ICI cancellation demodulation scheme is used. In this scheme, signal at the $(k+1)$ subcarrier is multiplied by "−1" and then added to the one at the $k$ subcarrier. Then, the resulting data sequence is used for making symbol decision.

### 2) ICI Cancelling Modulation

The ICI self-cancellation scheme requires that the transmitted signals be constrained such that $X(1) = -X(0), X(3) = -X(2), \ldots, X(N-1) = -X(N-2)$ using this assignment of transmitted symbols allows the received signal on subcarriers $k$ and $k+1$ to be written as

$$Y(k) = \sum_{l=0,2,4,6}^{N-2} X(l)[S(l-k) - S(l+1-k)] + n_k \qquad (3)$$

$$Y'(k+1) = \sum_{l=0,2,4,6}^{N-2} X(l)[S(l-k-1) - S(l-k)] + n_{k+1} \qquad (4)$$

and the ICI coefficient $S'(l-k)$ referred as

$$S'(l-k) = S(l-k) - S(l+1-k) \qquad (5)$$

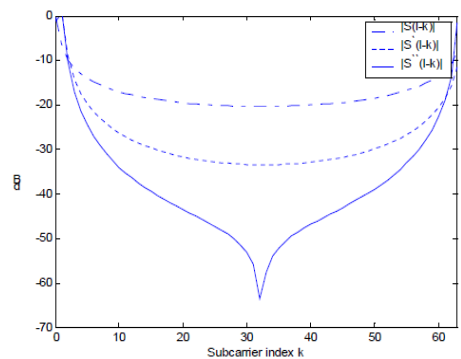

Fig 2: Comparison of |S(l-k)|, |S`(l-k)|, and |S``(l-k)| for N = 64 and ε = 0.4

Fig. 2 shows a comparison between |S'(l-k)| and |S(l-k)| on a logarithmic scale. It is seen that |S'(l-k)| << |S(l-k)| for most of the l-k values. Hence, the ICI components are much smaller than they are in |S(l-k)|. Also, the total number of interference signals is halved since only the even subcarriers are involved in the summation.

*3). ICI Canceling Demodulation*

ICI modulation introduces redundancy in the received signal since each pair of subcarriers transmit only one data symbol. This redundancy can be exploited to improve the system power performance, while it surely decreases the bandwidth efficiency. To take advantage of this redundancy, received signal at the $(k + 1)^{th}$ subcarrier, where k is even, is subtracted from the $k^{th}$ subcarrier.

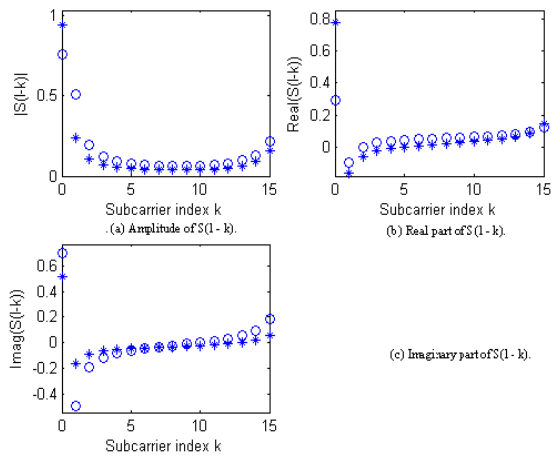

Fig. 3 An example of S(l - k) for N = 16; l = 0. (a) Amplitude of S(l - k). (b) Real part of S(l - k). (c) Imaginary part of S(l - k).

This is expressed mathematically as

$$Y''(k) = Y'(k) - Y'(k+1) =$$

$$\sum_{l=0}^{N-2} X(l)[-S(l-k-1)+2S(l-k)-S(l-k+1)]+n_k-n_k+1 \quad (6)$$

Subsequently, the ICI coefficients for this received signal becomes

$$S'(l-k) = -S(l-k-1)+2S(l-k)-S(l-k+1) \quad (7)$$

When compared to the two previous ICI coefficients $|S(1-k)|$ for the standard OFDM system and $|S(1-k)|$ for the ICI canceling modulation, $|S''(1-k)|$ has the smallest ICI coefficients, for the majority of l-k values, followed by $|S(1-k)|$ and $|S(1-k)|$. The combined modulation and demodulation method is called the ICI self-cancellation scheme. The reduction of the ICI signal levels in the ICI self-cancellation scheme leads to a higher CIR. The theoretical CIR can be derived as

$$CIR = \frac{|-S(-1)+2S(O)-S(1)|^2}{\sum_{l=2,4,6}^{N-1}|-S(l-1)+2S(l)-S(l+1)|^2} \quad (8)$$

As mentioned above, the redundancy in this scheme reduces the bandwidth efficiency by half. This could be compensated by transmitting signals of larger alphabet size. Using the theoretical results for the improvement of the CIR should increase the power efficiency in the system and gives better results for the BER. Hence, there is a tradeoff between bandwidth and power tradeoff in the ICI self-cancellation scheme. The Fig. 4 shows the model of the proposed method.

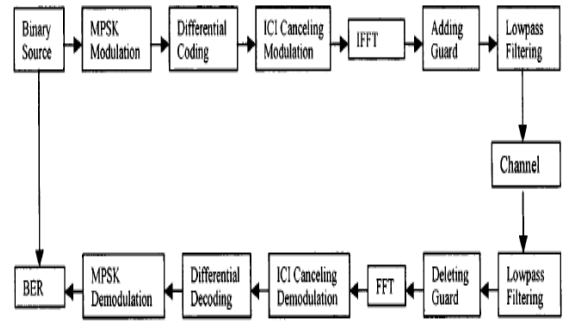

Fig.4. Simulation block diagram of the proposed system

ICI self-cancellation scheme can be combined with error correction coding. Such a system is robust to both AWGN and ICI, however, the bandwidth efficiency is reduced. The proposed scheme provides significant CIR improvement, which has been studied theoretically and by simulations. The scheme also works well in a multipath radio channel with Doppler frequency spread. Under the condition of the same bandwidth efficiency and larger frequency offsets, the proposed OFDM system using the ICI self-cancellation scheme performs much better than standard OFDM systems. In addition, since no channel equalization is needed for reducing ICI, the element without increasing system complexity.

Fig. 5 shows the comparison of the theoretical CIR curve of the ICI self-cancellation scheme, calculated by, and the CIR of a standard OFDM system is calculated. As expected, the CIR is greatly improved using the ICI self-cancellation scheme. The improvement can be greater than 15 dB for $0 < \varepsilon < 0.5$.

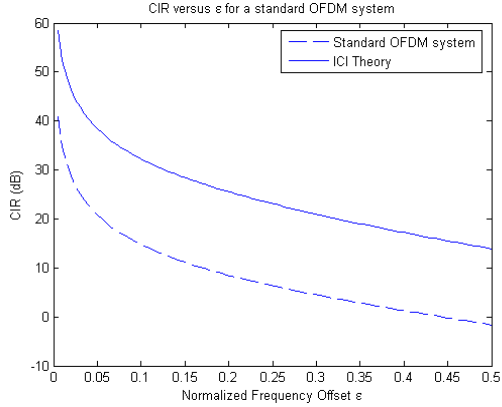

Fig. 5. CIR versus ε for a standard OFDM system

## IV. MAXIMUM LIKELIHOOD ESTIMATION

The second method for frequency offset correction in OFDM systems was suggested by Moose. In this approach, the frequency offset is first statistically estimated using a maximum likelihood algorithm and then cancelled at the receiver. This technique involves the replication of an OFDM symbol before transmission and comparison of the phases of each of the subcarriers between the successive symbols.

When an OFDM symbol of sequence length $N$ is replicated, the receiver receives, in the absence of noise, the $2N$ point sequence $\{r(n)\}$ is given by

$$\left| S''(1-k) \right| \qquad (11)$$

where $\{x(k)\}$ are the $2k+1$ complex modulation values used to modulate $2k+1$ subcarriers, $H(K)$ is the channel transfer function for $k^{th}$ carrier and ε is the normalized frequency offset of the channel.

### A. *Offset Estimation*

The first set of N symbols is demodulated using an $N$-point FFT to yield the sequence $R_1(k)$, and the second set is demodulated using another $N$-point FFT to yield the sequence $R_2(k)$. The frequency offset is the phase difference between $R_1(k)$ and $R_2(k)$, that is

$$R_2(k) = R_1(k) e^{j2\pi\varepsilon} \qquad (12)$$

adding the AWGN yields

$$Y_1(k) = R_1(k) + W_1(k) \qquad (13)$$

$$Y_2(k) = R_1(k) e^{j2\pi\varepsilon} + W_2(k) \qquad (14)$$

$$k = 0,1,....,N-1$$

This maximum likelihood estimate is a conditionally unbiased estimate of the frequency offset and was computed using the received data. The maximum likelihood estimate of the normalized frequency offset is given by:

$$\overset{\Lambda}{\varepsilon} = \frac{1}{2\pi} \tan^{-1} \left[ \frac{\sum\limits_{k=-K}^{K} \operatorname{Im} Y_2(k) Y_1^*(k)}{\sum\limits_{k=-K}^{K} \operatorname{Re} Y_2(k) Y_1^*(k)} \right] \qquad (15)$$

Once the frequency offset is known, the ICI distortion in the data symbols is reduced by multiplying the received symbols with a complex conjugate of the frequency shift and applying the FFT,

$$X(n) = FFT\{Y(n) e^{\frac{-j2\pi n\varepsilon}{N}}\} \qquad (16)$$

## V. EXTENDED KALMAN FILTERING

### A. *Problem Formulation*

A state-space model of the discrete Kalman filter is defined as

$$z(n) = a(n) d(n) + v(n) \qquad (17)$$

In this model, the observation z(n) has a linear relationship with the desired value d(n). By using the discrete Kalman filter, d(n) can be recursively estimated based on the observation of z(n) and the updated estimation in each recursion is optimum in the minimum mean square sense.
The received symbols are

$$Y(n) = X(n) e^{\frac{j2\pi n\varepsilon(n)}{N}} + W(n) \qquad (18)$$

It is obvious that the observation y(n) is in a nonlinear relationship with the desired value $\varepsilon_n$.
At the receiver

$$Y(n) = f(\varepsilon(n) + W(n) \qquad (19)$$

Where $\qquad f(\varepsilon(n) = X(n) e^{\frac{j2\pi n\varepsilon(n)}{N}} \qquad (20)$

In order to estimate ε efficiently in computation, we build an approximate linear relationship using the first-order Taylor's expansion:

$$y(n) \approx f\left(\varepsilon^{\wedge}(n-1)\right) + f'\left(\varepsilon^{\wedge}(n-1)\right)\left[\varepsilon(n) - \varepsilon^{\wedge}(n-1)\right] + w(n) \qquad (21)$$

where $\varepsilon^{\wedge}(n-1)$ is the estimation of $\varepsilon(n-1)$

$$f'\left(\in^{\wedge}(n-1)\right) = \frac{\partial f(\in(n))}{\partial \in(n)} \qquad (22)$$

$$= j\frac{2\pi n'}{N} e^{\frac{j2\pi n'\in(n-1)}{N}} \qquad (23)$$

and $\quad z(n) = y(n) - f\left(\varepsilon^{\wedge}(n-1)\right) \qquad (24)$

$$d(n) = \varepsilon(n) - \varepsilon^{\wedge}(n-1) \qquad (25)$$

and the following relationship:

$$z(n) = f'\left(\varepsilon(n-1)\right)d(n) + w(n) \qquad (26)$$

which has the same form as, i.e., z(n) is linearly related to d(n). Hence the normalized frequency offset $\varepsilon$ can be estimated in a recursive procedure similar to the discrete Kalman filter. As linear approximation is involved in the derivation, the filter is called the extended Kalman filter (EKF). The derivation of the EKF is omitted in this report for the sake of brevity.

*B. ICI Cancellation*

There are two stages in the EKF scheme to mitigate the ICI effect: the offset estimation scheme and the offset correction scheme.

*1). Offset Estimation Scheme*

To estimate the quantity $\varepsilon(n)$ using an EKF in each OFDM frame, the state equation is built as

$$\varepsilon(n) = \varepsilon(n-1) \qquad (27)$$

i.e., in this case we are estimating an unknown constant $\varepsilon$. This constant is distorted by a non-stationary process x(n), an observation of which is the preamble symbols preceding the data symbols in the frame. The observation equation is

$$Y(n) = X(n)e^{\frac{j2\pi n'\in(n)}{N}} + W(n) \qquad (28)$$

where y(n) denotes the received preamble symbols distorted in the channel, w(n) the AWGN, and x(n) the IFFT of the preambles X(k) that are transmitted, which are known at the receiver. Assume there are $N_p$ preambles preceding the data symbols in each frame are used as a training sequence and the variance $\sigma$ of the AWGN w(n) is stationary. The computation procedure is described as follows.

1. Initialize the estimate and corresponding state error P(0).

2. Compute the H(n), the derivative of y(n) with respect to $\varepsilon$ (n) at , the estimate obtained in the previous iteration.

3. Compute the time-varying Kalman gain K(n) using the error variance P(n- 1), H(n), and $\sigma^2$

4. Compute the estimate y^(n)using x(n) and )$\varepsilon^{\wedge}$(n-1)., i.e. based on the observations up to time n-1, compute the error between the true observation y(n) and y^(n).

5. Update the estimate $\varepsilon^{\wedge}(n)$ by adding the K(n)-weighted error between the observation y(n) and y^(n) to the previous estimation $\varepsilon^{\wedge}$(n-1).

6. Compute the state error P(n) with the Kalman gain K(n), H(n), and the previous error P(n-1).

7. If n is less than $N_p$, increment n by 1 and go to step 2; otherwise stop.

It is observed that the actual errors of the estimation $\varepsilon^{\wedge}$(n) from the ideal value $\varepsilon$(n) are computed in each step and are used for adjustment of estimation in the next step.

The pseudo code of computation is summarized as

Initialize P(n), $\varepsilon^{\wedge}$(0). For n=1,2....$N_p$ compute,

$$H(n) = \frac{\partial y(x)}{\partial x} \qquad (29)$$

$$= \frac{j2\pi n'}{N} e^{\frac{j2\pi n'\in(n-1)}{N}} X(n) \qquad (30)$$

$$K(n) = P(n-1)H*(n)[P(n-1) + \sigma^2]^{-1} \qquad (31)$$

$$\varepsilon^{\wedge}(n) = \varepsilon^{\wedge}(n-1) + \text{Re}\{K(n)[y(n) - x(n)e^{\frac{j2\pi n'\in(n-1)}{N}}]\} \qquad (32)$$

$$P(n) = \left[1 - K(n)H(n)\right]P(n-1) \qquad (33)$$

*2). Offset Correction Scheme*

The ICI distortion in the data symbols x(n) that follow the training sequence can then be mitigated by multiplying the received data symbols y(n) with a complex conjugate of the estimated frequency offset and applying FFT, i.e.

$$X^{\wedge}(n) = FFT\{Y(n)e^{\frac{j2\pi n'\in}{N}}\} \qquad (34)$$

As the estimation of the frequency offset by the EKF scheme is pretty efficient and accurate, it is expected that the performance will be mainly influenced by the variation of the AWGN.

## VI. SIMULATED RESULT ANALYSIS

### A. Performance

In order to compare the three different cancellation schemes, BER curves were used to evaluate the performance of each scheme. For the simulations in this paper, MATLAB was employed with its Communications Toolbox for all data runs. The OFDM transceiver system was implemented as specified by Fig. 1. Frequency offset was introduced as the phase rotation. Modulation schemes of binary phase shift keying (BPSK) and Quadrature amplitude modulation (QAM) were chosen as they are used in many standards such as 802.11a. Simulations for cases of normalized frequency offsets equal to 0.05, 0.15, and 0.30.

Table 6.1: Simulation Parameters

| PARAMETERS | VALUES |
|---|---|
| Number of carriers | 768 |
| Modulation | BPSK,QAM |
| Frequency offset | [0,0.15,0.30] |
| No. of OFDM symbols | 100 |
| Bits per OFDM symbols | N*log2(M) |
| $E_b$-$N_o$ | 1:15 |
| IFFT size | 1024 |

### B. BER Performance

Fig. 6 to Fig.11 provides comparisons of the performance of the SC, ML and EKF schemes for different alphabet sizes and different values of the frequency offset.

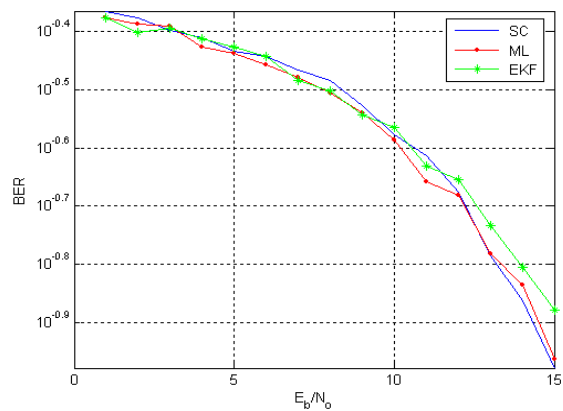

Fig. 6. BER Performance with ICI Cancellation, ε=0.15 for 4-BPSK

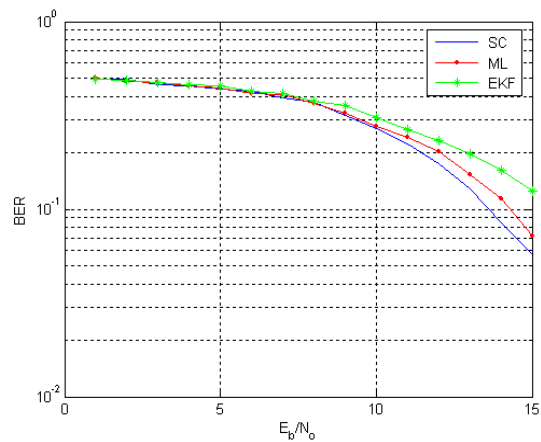

Fig. 7: BER Performance with ICI Cancellation, ε=0.30 for 16-BPSK

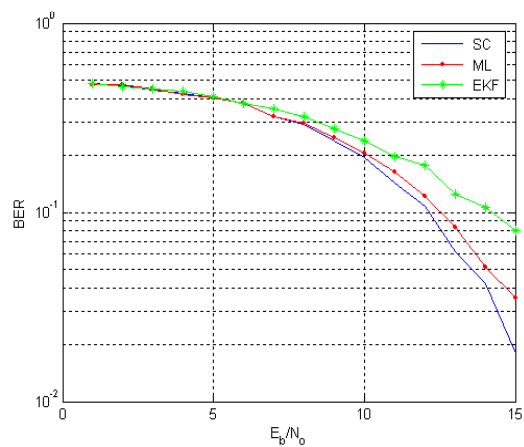

Fig. 8: BER Performance with ICI Cancellation, ε=0.05 for 64-BPSK

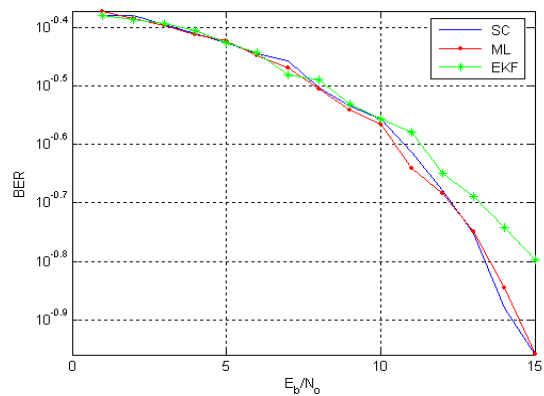

Fig. 9: BER Performance with ICI Cancellation   ε=0.15 for 4-OAM

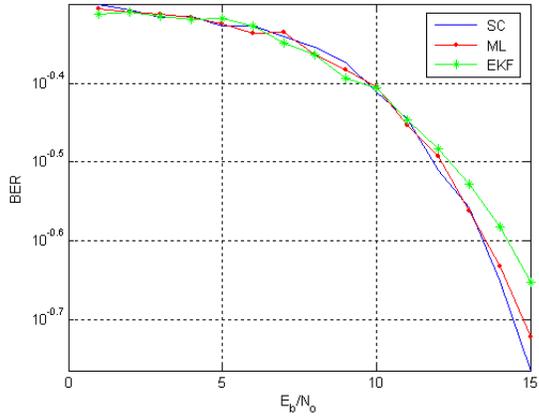

Fig. 10: BER Performance with ICI Cancellation, ε=0.30 for 16-QAM

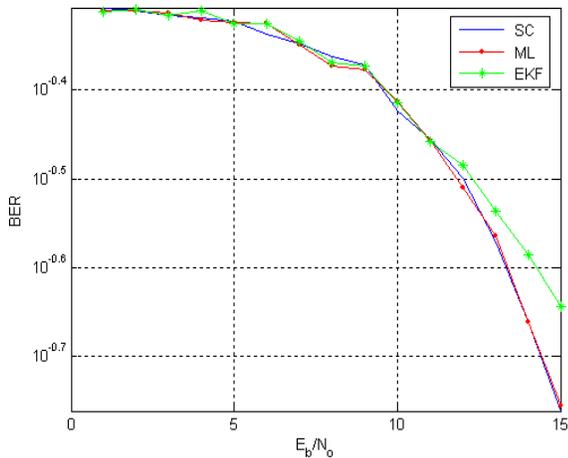

Fig. 11. BER Performance with ICI Cancellation, ε=0.05 for 64-QAM

It is observed in the figures that each method has its own advantages. In the presence of small frequency offset and binary alphabet size, self cancellation gives the best results. However, for larger alphabet sizes and larger frequency offset such as 4-BPSK and frequency offset of 0.30, self cancellation does not offer much increase in performance. The maximum likelihood method gives the best overall results. The Kalman filter method indicates that for very small frequency offset, it does not perform very well, as it hardly improves BER. However, for high frequency offset the Kalman filter does perform extremely well. It gives a significant boost to performance. Significant gains in performance can be achieved using the ML and EKF methods for a large frequency offset.

These results also show that degradation of performance increases with frequency offset. For the case of BPSK, even severe frequency offset of 0.30 does not deteriorate the performance too greatly. However, for QAM with an alphabet of size 2, performance degrades more quickly. When frequency offset is small, the 4-QAM system has a lower BER than the BPSK system. But the BER of 4-QAM varies more dramatically with the increase the frequency offset than that of BPSK. Therefore it is concluded that larger alphabet sizes are more sensitive to ICI. Tables 6.2 and 6.3 summarize required values of SNR for BER specified at $10^{-6}$

Table 6.2 Required SNR and improvement for BER of 10^-6 for BPSK

| Sl No. | Method | ε= 0.05 | ε= 0.15 | ε= 0.30 |
|---|---|---|---|---|
| 1 | SC | 12 dB | 11dB | 13dB |
| 2 | ML | 12.5 dB | 12 dB | 12 dB |
| 3 | EKF | 13 dB | 12dB | 13.5 dB |

Table 6.3 Required SNR and improvement for BER of 10^-6 for QAM

| Sl. No. | Method | ε= 0.05 | ε= 0.15 | ε= 0.30 |
|---|---|---|---|---|
| 1 | SC | 13 dB | 12 dB | 11 dB |
| 2 | ML | 11 dB | 12 dB | 13dB |
| 3 | EKF | 12dB | 13 dB | 14 dB |

For small alphabet sizes and for low frequency offset values, the SC and ML techniques have good performance in terms of BER. However, for higher order modulation schemes, the EKF and ML techniques perform better. This is attributed to the fact that the ML and EKF methods estimate the frequency offset very accurately and cancel the offset using this estimated value. However, the self-cancellation technique does not completely cancel the ICI from adjacent sub-carriers, and the effect of this residual ICI increases for larger alphabet sizes and offset values.

## VII. CONCLUSION

In this paper, the performance of OFDM systems in the presence of frequency offset between the transmitter and the receiver has been studied in terms of the Carrier-to-Interference ratio (CIR) and the bit error rate (BER) performance. Inter-carrier interference (ICI) which results from the frequency offset degrades the performance of the OFDM system.

Three methods like ICI self-cancellation (SC), maximum likelihood (ML) estimation and Kalman filtering (EKF) methods were explored in this paper for mitigation of the ICI. The cancellation of the frequency offset has been investigated in this paper and compared with these three techniques. The choice of which method to employ depends on the specific application. For example, self cancellation does not require very complex hardware or software for implementation. However, it is not bandwidth efficient as there is a redundancy of bits for each carrier. The ML method also introduces the same level of redundancy but provides better BER performance, since it accurately estimates the frequency offset. Its implementation is more complex than the SC method. On the other hand, the EKF method does not reduce bandwidth efficiency as the frequency offset can be estimated from the preamble of the data sequence in each OFDM frame.

However, it has the most complex implementation of the three methods. In addition, this method requires a training sequence to be sent before the data symbols for estimation of the frequency offset. It can be adopted for the receiver design for IEEE 802.11a because this standard specifies preambles for every OFDM frame. The preambles are used as the training sequence for estimation of the frequency offset.

In this paper, the simulations were performed in an AWGN channel. This model can be easily adapted to a flat-fading channel with perfect channel estimation. Further work can be done by performing simulations to investigate the performance of these ICI cancellation schemes in multipath fading channels without perfect channel information at the receiver. In this case, the multipath fading may encumber the performance of these ICI cancellation schemes.

### AUTHORS PROFILE

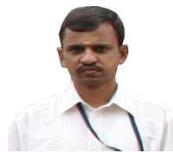

B.Sathish Kumar currently is a Senior Lecturer, Sri Ramakrishna Engineering College, Coimbatore, Tamil Nadu, India. He received the Masters Degree from Anna University Chennai, in the year 2005 and currently pursuing PhD in Anna University Coimbatore. His research interest includes Wireless Communication, Networking, Signal Processing, Mobile Communication and Multicarrier Communication.

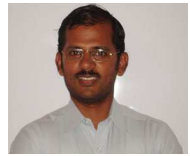

K.R.Shankar Kumar currently is a Professor, Sri Ramakrishna Engineering College, Coimbatore, Tamil Nadu, India. He received the Masters Degree from Madras University, in the year 2000 and the PhD from Indian Institute of Science, Bangalore, in the year 2004. His research interest includes future broad band wireless communication, Multicarrier Communication, Digital Communication, Advanced Signal Processing for communication. He has published more than 20 Journal papers in the field of CDMA systems. His research work was supported by Swarnajayanti Fellowship, Department of Science and Technology (DST), Government of India.

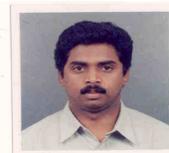

Radhakrishnan Rathinavel is currently Professor, Sri Ramakrishna Engineering College, in Electronics and Communication Engineering Department, Coimbatore, Tamil Nadu, India. He received his Masters Degree from P.S.G.College of Technology, Coimbatore, in the year 1997 and the PhD from Anna University Chennai in the year 2008. His research interest includes Wireless Communication, Signal Processing, Networking and Mobile Communication. He has published more than 11 Journal papers in the field of CDMA systems, Mobile communication, Wireless Networking and Signal Processing.